\documentclass[conference]{IEEEtran}
\IEEEoverridecommandlockouts

\usepackage{cite}

\usepackage{amsfonts}
\usepackage{algorithmic}
\usepackage{textcomp}
\usepackage{xcolor}
\def\BibTeX{{\rm B\kern-.05em{\sc i\kern-.025em b}\kern-.08em
    T\kern-.1667em\lower.7ex\hbox{E}\kern-.125emX}}

\newcommand{\mobilenetvtwo}{MobileNetV2}

\usepackage{amsmath} 

\usepackage{textcomp}
\usepackage{xcolor}

\usepackage{graphics, epstopdf, graphicx, epsfig, psfrag, epsf,subfigure}

\usepackage{algorithm}
\usepackage{algorithmic}

\usepackage{url}

\newtheorem{example}{Example}

\newcommand{\ie}{{\em i.e., }}
\newcommand{\eg}{{\em e.g., }}

\newcommand{\Tau}{\mathcal{T}}

\newcommand{\Nset}{\mathcal{N}}

\newcommand{\Iset}{\mathcal{I}}
\newcommand{\Lset}{\mathcal{L}}

\newcommand{\Oset}{\mathcal{O}}

\begin{document}
%\ninept

\IEEEoverridecommandlockouts
\IEEEpubid{\makebox[\columnwidth]{979-8-3503-5209-2/24/\$31.00~\copyright 2024 IEEE \hfill}
\hspace{\columnsep}\makebox[\columnwidth]{ }}

\title{Early-Exit meets Model-Distributed Inference at Edge Networks
}

\author{\IEEEauthorblockN{
Marco Colocrese, Erdem Koyuncu and Hulya Seferoglu}
\IEEEauthorblockA{\small mcoloc2@uic.edu,  ekoyuncu@uic.edu, hulya@uic.edu}
\IEEEauthorblockA{University of Illinois Chicago
}
\thanks{This work was supported in parts by the Army Research Lab (\#W911NF2120272), the Army Research Office (\#W911NF2410049), and the National Science Foundation (CCF-1942878, CNS-2148182, CNS-2112471).}
}

\maketitle

\begin{abstract}
Distributed inference techniques can be broadly classified into data-distributed and model-distributed schemes. In data-distributed inference (DDI), each worker carries the entire deep neural network (DNN) model but processes only a subset of the data. However, feeding the data to workers results in high communication costs, especially when the data is large. An emerging paradigm is model-distributed inference (MDI), where each worker carries only a subset of DNN layers. In MDI, a source device that has data processes a few layers of DNN and sends the output to a neighboring device, \ie offloads the rest of the layers. This process ends when all layers are processed in a distributed manner. In this paper, we investigate the design and development of MDI with early-exit, which advocates that there is no need to process all the layers of a model for some data to reach the desired accuracy, \ie we can exit the model without processing all the layers if target accuracy is reached. We design a framework MDI-Exit that adaptively determines early-exit and offloading policies as well as data admission at the source.  Experimental results on a real-life testbed of NVIDIA Nano edge devices show that MDI-Exit processes more data when accuracy is fixed and results in higher accuracy for the fixed data rate.

\end{abstract}

\section{Introduction}
\label{sec:intro}
The traditional approach to performing deep neural network (DNN) inference in a distributed fashion is data-distributed inference (DDI). A typical DDI scenario is when an end user or edge server, \ie a source device, would like to classify its available data. Workers may correspond to other end users, edge servers, or remote cloud. The source device itself could function as one of the workers by processing some of its own data.  The source partitions and transmits the available data to multiple workers, all of which carry the same DNN model. The workers then perform inference on their received data via the available DNN. The outputs are finally sent back to the source. This strikingly simple approach, although very effective in certain cases, incurs very high communication costs especially when the input data size is large.

An alternative to DDI is model-distributed inference (MDI), which is often referred to as model parallelism \cite{AR-MDI}. In this scenario, the DNN model itself is partitioned into blocks of layers and then distributed across multiple workers. Unlike DDI, the data is not distributed and resides at the source. Instead, given an input to the DNN, the source may process the input through the first few layers of the DNN. The resulting feature vectors (intermediate outputs of the DNN after the first few layers) are then passed to the next worker that is responsible for the next few subsequent layers. The inference is completed sequentially in the same manner by passing feature vectors to the workers responsible for processing them. After the output of the DNN is finally computed, it is transmitted back to the source. Processing at the workers is done in parallel to take advantage of pipelining \cite{AR-MDI}.

\begin{figure}[t!]
%\vspace{-5pt}
    \centering
    \scalebox{0.39}{
    \includegraphics{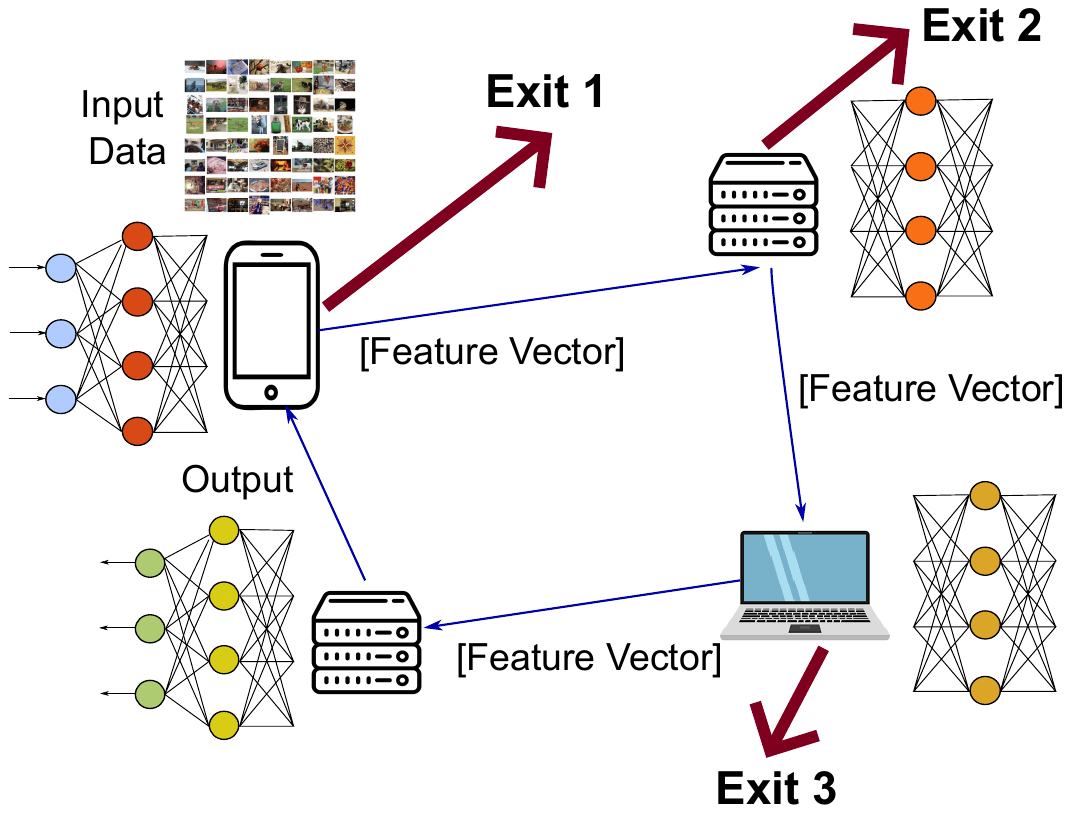}}
    \vspace{-180pt}
    \caption{Model-distributed inference with early exit. }
    \label{fig:MDI-Exit}
    \vspace{-10pt}
\end{figure}

The performance of MDI with heterogeneous resources is investigated \cite{AR-MDI} and an adaptive layer allocation mechanism across workers is designed. It is shown that MDI significantly reduces the inference time as compared to data distributed inference when the size of data is large \cite{AR-MDI}. MDI is applied to a multi-source scenario in \cite{li2023model}. The goal of this paper is to further improve the performance of MDI, \ie reduce the inference time without hurting accuracy by exploiting early-exit mechanisms in MDI.

As in most deep learning applications, in the context of computer vision tasks, specifically classification, it is important to acknowledge that not all inputs exhibit the same level of difficulty, which pertains to the complexity of their respective classifications. When designing and training a DNN, the primary objective is to maximize accuracy, thereby creating systems capable of accurately classifying more complex inputs. However, even within the same dataset, certain inputs can be easily classified with fewer computations. This concept underlies the notion of ``network overthinking'' introduced in \cite{shallowdeep}. The authors propose that a network engages in overthinking when a previous layer's representation is sufficient for accurate classification.
To exploit this property of certain inputs, early-exit can be used, reducing inference delay. Early-exit, first introduced in \cite{branchynet},  allows early termination of inference, providing a trade-off between accuracy and computation cost. It enables DNN models to make predictions at multiple points during the computation process, allowing for early classification if a confident prediction is obtained. This approach is particularly useful in scenarios where computational resources are limited or real-time processing is required. The goal of this paper is to take advantage of early exit mechanisms in MDI as illustrated in the next example. 

\begin{example}
    Consider a scenario in Fig. \ref{fig:MDI-Exit}, where four workers are connected to each other. Worker 1 has the input data (images) and want to classify them using a pre-trained DNN model. According to MDI, the model is partitioned and distributed across workers \cite{li2023model}.  Workers process the partial models (layers) that are assigned to them. The final output is transmitted back to Worker 1 (since it has data and is interested in the output). With early-exit, Worker 1 processes a few layers and passes the feature vector to a classification layer. Then, it calculates its confidence level (which is detailed in Section \ref{sec:system}) of classification. Early-exit advocates that if the calculated confidence level is larger than a threshold, the inference process can be terminated. Worker 1 can terminate the inference process without sending its feature vector to the next worker if the threshold is reached. Otherwise, Worker 1 does not early exit, and it sends its feature vector to Worker 2 for processing later layers. Similarly, each worker explores early-exit opportunities (\ie if the threshold is reached). If a worker reaches the confidence level threshold, it exits and sends the output to Worker 1. In this setup, each worker needs to decide on (i) which layers to process, (ii) whether to exit early, and (iii) which worker to send its feature vector. 
\end{example}

In this paper, we design a model-distributed inference with early-exit (MDI-Exit), which adaptively and in a decentralized manner determines the model allocation and early-exit decisions at each worker. Our approach stems from Network Utility Maximization (NUM) \cite{neely2010stochastic}, where model allocation and early-exit decisions are made based on queue sizes. We implement MDI-Exit in a real testbed consisting of NVIDIA Jetson TX2s. The experimental results show that MDI-Exit processes more data when accuracy is fixed and results in higher accuracy when the data rate is fixed as compared to baselines for MobileNetV2 \cite{sandler2018mobilenetv2} and ImageNet \cite{deng2009imagenet}.

The rest of the paper is organized as follows. In Section \ref{sec:related}, we present the related work. We introduce our system model and provide background in Section \ref{sec:system}. Section \ref{sec:MDI-Exit} presents our MDI-Exit design. In Section \ref{sec:results}, we provide experimental results. We draw our main conclusions in Section \ref{sec:conc}.

\section{Related Work}
\label{sec:related}

A defining characteristic of edge computing networks is the severe limitations of the constituent nodes in terms of their communication bandwidth, battery/power limitations, computing/transmission power, etc. Such limitations make centralized approaches to learning or inference infeasible. Hence, developing novel distributed learning and inference methods is crucial, especially at the edge.

An earlier work on distributed inference is a two-part DNN partition \cite{teerapittayanon2017distributed}. Specifically, a DNN is split into two parts, where the first few layers are placed at the end user, while the remaining layers are located in the cloud. The processing in the end user and cloud could be iterative or parallel. In iterative processing, an end user processes a few layers, and if it is confident with the output, it terminates the processing. Otherwise, feature vectors are sent to the cloud for further processing at subsequent DNN layers. In parallel processing, layers are partitioned and assigned to the end user and edge server (or cloud), which operate in parallel \cite{kang2017neurosurgeon} and \cite{eshratifar2019jointdnn}. In order to take full advantage of edge networks, model distributed inference with multiple split points is developed. A model partitioning problem for MDI is formulated in \cite{EdgePipe-hu2021pipeline} using dynamic programming. An adaptive and resilient layer allocation mechanism is designed in \cite{AR-MDI} to deal with the heterogeneous and time-varying nature of resources. Model-distributed inference for multiple sources is developed in \cite{li2023model}. As compared to this line of work, we focus on merging model-distributed inference with early-exit.

In many real-world datasets, certain data inputs may consist of much simpler features as compared to other inputs \cite{bolukbasi}. In such a scenario, it becomes desirable to design more efficient architectures that can exploit the heterogeneous complexity of dataset members.  This can be achieved by introducing additional internal classifiers and exit points to the models \cite{biasielli2020neural,panda2016cdl, branchynet, shallowdeep, koyuncu:c31}. These exit points prevent simple inputs from traversing the entire network, reducing the computational cost.

We note that various other methods to reduce the memory, communication, and computation footprints of DNNs have been proposed, including conditional architectures \cite{han2021dynamic},  pruning \cite{liu2020pruning,koyuncu:c32}, quantization \cite{chen2015compressing,gong2014compressing}, and gradient sparsification \cite{wangni2018gradient, alistarh2018convergence}.  
Our work is complementary to these alternate techniques in the sense that the performance of MDI-Exit can be further improved using one or more of these methods.

\section{System Model and Background}
\label{sec:system}
{\bf Setup.} We consider an edge network with end users and edge servers. We name these devices as workers. There are $N$ workers in the system, and the set of workers is $\Nset$. We name the $n$th worker as ``Worker $n$''. 
One of these workers is the source, which already has or collects data (images) in real time from the environment. The source worker would like to classify its data as soon as possible by exploiting its own and other workers' computing power.

We consider a dynamic edge computing setup where workers join and leave the system anytime. Also, computing and communication delay of workers could be potentially heterogeneous and vary over time.  Workers form an ad-hoc topology depending on their geographical locations, proximity to each other, and surrounding obstacles.

\textbf{DataSet and DNN Model.} Suppose that Worker $n$ is a source. It may already have data or collect data from the environment in real time. The $d$th vectorized data is $A_d \in \mathcal{R}^u$.

We assume a pre-trained DNN model of $L$ layers is used. The set of layers is determined by $\Lset = \{ l_1, \ldots, l_L \}$, where $l_l$ is the $l$th layer. The feature vector at the output of the $l$th layer is $a_l(d)$ for data $A_d$. The output of the DNN model is $y \in \mathcal{R}^v$. %, where $v$ equals to the number of labels.  

\textbf{Early-Exit.} We consider that that there are $K$ exit points in an $L$-layer DNN model, where $K < L$. Assume that the $k$th exit point is after the $l$th layer. The feature vector $a_l(d)$ produced by the $l$th layer for data $A_d$ is fed to a classifier for the $k$th exit point according to early-exit. The classifier produces vector $b_l^k(d)$ with size equal to $v$. In other words, $b_l^k(d) = [x_1^k(d), \ldots x_v^k(d)]$. Following the same idea in \cite{koyuncu:c31}, softmax is used after the classification layer, which normalizes the output of the classification layer according to 
\begin{align}
    c_i^k(d) = \frac{e^{x_i^k(d)}}{\sum_{\nu=1}^{v} e^{x_{\nu}^k(d)} }, \forall i \in \{1, \ldots v\}.
\end{align} The confidence level $C_k(d)$  at the $k$th exit point for data $A_d$ is expressed as 
\begin{align} \label{eq:Confidence-k}
    C_k(d) = \max_{i \in \{1, \ldots, v\}} \{c_i^k(d)\}.
\end{align}  If the confidence level $C_k(d)$ is larger than an early-exit threshold $T_e^k$ for the $k$th exit point, early-exit happens. 
%and $T_e^k$ is the early-exit threshold for the $k$th exit point. 

\textbf{Model Partitioning.} We partition the model at early-exit points. The DNN layers that are between $k-1$th and $k$th exit points are called task $k$ and are represented by $\tau_k(d)$ for data $d$. Considering that the first layer in task $k$ is $\lambda_b^k \in \Lset$ and the last layer is  $\lambda_e^k \in \Lset$, task $\tau_k(d)$ is defined as processing all the layers in $\Lset_k = \{ \lambda_b^k, \lambda_b^k + 1, \ldots, \lambda_e^k \}$ with input feature vector $a_{\lambda_b^k}(d)$. Since there are $K$ exit points in the model (including the actual output), the set of tasks is $\Tau = \{\tau_1(d), \ldots, \tau_K(d)\}$ for data $A_d$. If there is no early exit point in a model, then there is only one task, and this task corresponds to processing all the layers in the model, \ie  $\Lset = \{ l_1, \ldots, l_L \}$.

\textbf{Queues.} Each worker $n$ maintains two queues: (i) an input queue $\Iset_n$, which keeps the tasks that Worker $n$ is supposed to process, and (ii) an output queue $\Oset_n$, which keeps the tasks that will be offloaded to other workers. The number of tasks in input and output queues are $I_n$ and $O_n$, respectively. 
If Worker $n$ has task $\tau_k(d)$ in its input queue $\Iset_n$ for data $A_d$ and needs to process this task, it identifies and processes the layers corresponding to this task; \ie $\Lset_k$ with input feature vector $a_{\lambda_b^k}(d)$. The MDI process can be pipelined so that Worker $n$ can start processing the next task in its input queue immediately after processing a task, \eg processing $\tau_k(d+1)$ immediately after $\tau_k(d)$.

\section{Model-Distributed Inference with Early-Exit}
\label{sec:MDI-Exit}

In this section, we present our model-distributed inference with early exit (MDI-Exit) design. We first present inference, early-exit, and offloading policies of MDI-Exit. Then, we detail data admission policies of MDI-Exit. 

\subsection{Inference, Early-Exit, and Offloading}

\textbf{Inference and Early-Exit.} The first step of our MDI-Exit design is to determine when to perform inference and early-exit. Our policy is summarized in Alg. \ref{alg:Inf-Exit}. In particular, Worker $n$ checks its input queue $\Iset_n$, which stores the tasks that are supposed to be processed by Worker $n$. If there exists a task in $\Iset_n$, let us assume that this task is $\tau_k(d)$, Worker $n$ performs inference task by processing the layers in $\Lset_k$ corresponding to the task $\tau_k(d)$. After processing all the layers in  $\Lset_k$, Worker $n$ reaches the $k$th exit point. Therefore, it checks if early-exit is possible or not. The feature vector at the exit point is passed through a classifier as explained in Section \ref{sec:system} and the confidence level $C_k(d)$ is calculated according to (\ref{eq:Confidence-k}). If the confidence level $C_k(d)$ is larger than the early-exit threshold $T_e^k$, then Worker $n$ exists from the DNN model for data $A_d$ and sends the output of the classifier to the source. On the other hand, \ie if the confidence level is not reached, Worker $n$ creates a new task $\tau_{k+1}(d)$ for data $A_d$ until the $k+1$th early-exit point. Worker $n$ determines which queue the task $\tau_{k+1}(d)$ should be put according to the sizes of input and output queues. In particular, if the input queue $\Iset_n$ is empty or the size of the output queue; $O_n$ is larger than a predetermined threshold $T_O$, then the new task $\tau_{k+1}(d)$ is inserted in the input queue as it means that local processing of tasks at Worker $n$ is faster. Otherwise, it is inserted in the output queue as it means that offloading is faster.

\begin{algorithm}[t!]
\caption{Inference and Early-Exit of MDI-Exit at Worker $n$.}\label{alg:Inf-Exit}
\begin{algorithmic}[1]
\STATE \textbf{Input:}  Early-exit threshold $T_e^k$, $\forall k$. Output queue threshold $T_O$. Input queue $\Iset_n$. Output queue $\Oset_n$.
\WHILE{There exists a task in $\Iset_n$}
\STATE Process the head-of-line task $\tau_k(d)$.
\STATE Calculate the confidence level  $C^k(d)$ according to (\ref{eq:Confidence-k}). 
\IF{$C^k(d) > T_e^k$}
\STATE Early exit and send the output of the classifier ($b_l^k(d)$) to the source. 
\ELSE
\IF{The input queue $\Iset_n$ is empty OR $O_n > T_O$ }
\STATE Insert the new task $\tau_{k+1}(d)$ to $\Iset_n$.
\ELSE
\STATE Insert the new task to $\Oset_n$.
\ENDIF
\ENDIF
\ENDWHILE
\end{algorithmic}
\end{algorithm}

\textbf{Offloading.} The next step in our MDI-Exit design is offloading. Worker $n$ offloads the tasks in its output queue to its one-hop neighbors. The task offloading policy from Worker $n$ to Worker $m$, considering that they are one-hop neighbors, is summarized in Alg. \ref{alg:Offloading} and will be detailed next. 

If there exists a task in the output queue $\Oset_n$, Worker $n$ explores the opportunities of offloading its tasks. Worker $n$ periodically learns from its one-hop neighbor Worker $m$ its input queue size $I_m$, per task computing delay $\Gamma_m$\footnote{We note that exit points could be arranged in a way that each task has a similar amount of computation requirements. Thus, $\Gamma_m$ can be measured independently from the identities of the tasks.} as well as transmission delay $D_{nm}$ between Workers $n$ and $m$. Worker $n$ runs Alg. \ref{alg:Offloading} for each of its one-hop neighbors. If the output queue size of Worker $n$ is larger than the input queue size of Queue $m$, \ie $O_n > I_m$ is satisfied; Worker $n$ concludes that its neighbor Worker $m$ is in a better state than itself in terms of processing tasks. Next, Worker $n$ compares the task completion delays. 

If a task at Worker $n$ is processed locally, it needs to wait so that all the tasks in the input queue are processed, so its waiting time is $I_n\Gamma_n$. If it is offloaded to Worker $m$, it needs to be transmitted and wait for all the tasks in the input queue of Worker $m$, so its waiting time is $D_{nm} + I_m \Gamma_m$. Thus, Worker $n$ compares these delays. If $I_n\Gamma_n > D_{nm} + I_m \Gamma_m$, the task is offloaded to Worker $m$ (line 3). Otherwise, the task is offloaded to Worker $m$ with  probability $\min\{\frac{I_n\Gamma_n }{D_{nm} + I_m\Gamma_m}, 1\}$ (line 5). The reason behind the probabilistic offloading is to utilize resources effectively. Even if offloading a task introduces a higher delay, offloading may still 
be good, considering that $O_n > I_m$ and the probabilistic transmission occur when local processing and offloading delays are on similar orders.

\begin{algorithm}[t!]
\caption{Offloading from Worker $n$ to $m$.}\label{alg:Offloading}
\begin{algorithmic}[1]
\WHILE{There exists a task in $\Oset_n$}
\IF{$O_n > I_m$ AND $I_n\Gamma_n  >  D_{n,m} + I_m \Gamma_m$} 
\STATE Offload the head-of-line task from $\Oset_n$ to Worker $m$.
\ELSIF{$O_n > I_m$}
\STATE Offload the head-of-line task from $\Oset_n$ to Worker $m$ with probability $\min\{\frac{I_n\Gamma_n }{D_{nm} + I_m\Gamma_m}, 1\}$.
\ENDIF
\ENDWHILE
\end{algorithmic}
\end{algorithm}

\subsection{Data Admission Policies} Suppose that Worker $n$ is the source. It may already have the data or collect the data from the environment according to some distribution. The question in this context is how to guarantee that all the data at the source is admitted and processed for DNN inference. We consider two scenarios: (i) Data arrival rate is adapted assuming that early-exit threshold is fixed. (ii) Early-exit threshold is adapted assuming that data arrival follows a general distribution.

\textbf{Data Arrival Rate Adaptation.} We first fix the early-exit confidence threshold $T_e^k$ for all tasks. Our goal is to adapt the arrival rate of the data that is collected/generated at the source. In other words, the source worker collects/generates data by following the ``inference rate'' that MDI-Exit imposes.  This scenario suits well for applications where a certain inference accuracy (confidence threshold) is needed while data arrival rate can be adapted. 

Our policy determines the interarrival time of data $\mu$ at the source. Our approach is inspired by congestion control mechanisms of delay-based TCP algorithms, \eg TCP Vegas. Let us assume that the source worker is Worker $n$. Our algorithm summarized in Alg. \ref{alg:interarrivalTime} uses the queue size information to determine the interarrival time $\mu$ at the source. In particular, if $I_n + O_n < T_{Q1}$, then the interarrival time $\mu$ is reduced according to line 3 to increase the data rate, where $\alpha$ is a predetermined constant between $0$ and $1$. The data rate is increased (interarrival time is reduced) in this case, because the number of tasks in input and output queues is very low, which means that the MDI-Exit system can handle more data. If $I_n + O_n > T_{Q1}$ but $I_n + O_n < T_{Q2}$, interarrival time is reduced according to line 5, but not as dramatically as line 3, because $\alpha > \beta$. Finally, if $I_n + O_n > T_{Q2}$, interarrival time is increased according to line 7 to decrease the data rate because queues are highly occupied. The data rate should be reduced so that the queue sizes remain bounded. We note that after every update on interarrival time $\mu$, the algorithm sleeps for $s$ seconds to give some time to queues to update their states. Alg. \ref{alg:interarrivalTime} runs again after the sleep time is over.

\begin{algorithm}[t!]
\caption{Data Interarrival Time Adaptation at Source.}\label{alg:interarrivalTime}
\begin{algorithmic}[1]
\STATE \textbf{Input:} Assume source worker is Worker $n$. Input queue size at the source $I_n$. Output queue size at the source $O_n$. Constants $\alpha$, $\beta$, and $\zeta$ between 0 and 1 and $\alpha > \beta$. Queue thresholds $T_{Q1}$ and $T_{Q2}$, where $T_{Q1} \leq T_{Q2}$. Sleep duration $s$. 
\IF{$I_n+O_n < T_{Q1}$}
\STATE $\mu = \mu - \alpha \mu$. Sleep for $s$ seconds. 
\ELSIF{$I_n+O_n > T_{Q1}$ AND $I_n+O_n < T_{Q2}$}
\STATE $\mu = \mu - \beta \mu$. Sleep for $s$ seconds. 
\ELSIF{$I_n+O_n > T_{Q2}$}
\STATE $\mu = \mu + \zeta \mu$. Sleep for $s$ seconds. 
\ENDIF
\end{algorithmic}
\end{algorithm}

\textbf{Early-Exit Threshold Adaptation.} In this setup, we consider that all the traffic arriving to the source worker should be processed, which can only be feasible with reduced accuracy. In other words, if the data rate is too high, more early-exits should happen (lower early-exit threshold, hence lower accuracy) to accommodate all the data. Otherwise, early-exit threshold (hence accuracy) can be increased as less data needs to be processed. Our approach, summarized in Alg. \ref{alg:ConfLevel}, follows a similar idea in Alg. \ref{alg:interarrivalTime}. In particular, if queues are not heavily occupied, the early-exit threshold is increased (lines 3 and 5). We note that the largest value of the early-exit threshold could be 1 due to the normalization in (\ref{eq:Confidence-k}), so we use $T_e = \min\{1, T_e + \alpha T_e\}$ and $T_e = \min\{1, T_e + \beta T_e\}$ in lines 3 and 5 respectively. On the the other hand, if queue occupancy is high (line 6), then the early-exit threshold is reduced (line 7), noting that early-exit threshold cannot be less than the minimum early-exit threshold $T_e^{\min}$ that we set.

\begin{algorithm}[t!]
\caption{Confidence Level Adaptation at Worker $n$.}\label{alg:ConfLevel}
\begin{algorithmic}[1]
\STATE \textbf{Input:} Input queue size  $I_n$. Output queue size $O_n$. Constants $\alpha$, $\beta$, and $\zeta$ between 0 and 1 and $\alpha > \beta$. Queue thresholds $T_{Q1}$ and $T_{Q2}$, where $T_{Q1} \leq T_{Q2}$. Sleep duration $s$. Current confidence level $T_e$.  Minimum confidence level $T_e^{\min} > 0$.
\IF{$I_n+O_n < T_{Q1}$}
\STATE $T_e = \min\{1, T_e + \alpha T_e\}$. Sleep for $s$ seconds. 
\ELSIF{$I_n+O_n > T_{Q1}$ AND $I_n+O_n < T_{Q2}$}
\STATE $T_e = \min\{1, T_e + \beta T_e\}$. Sleep for $s$ seconds. 
\ELSIF{$I_n+O_n > T_{Q2}$}
\STATE $T_e = \max\{T_e^{\min}, T_e - \zeta T_e\}$. Sleep for $s$ seconds. 
\ENDIF
\STATE $T_e^k  \leftarrow T$, $\forall k$. 
\end{algorithmic}
\end{algorithm}

\begin{figure}[t]
    \centering
    \includegraphics[width=9.5cm]{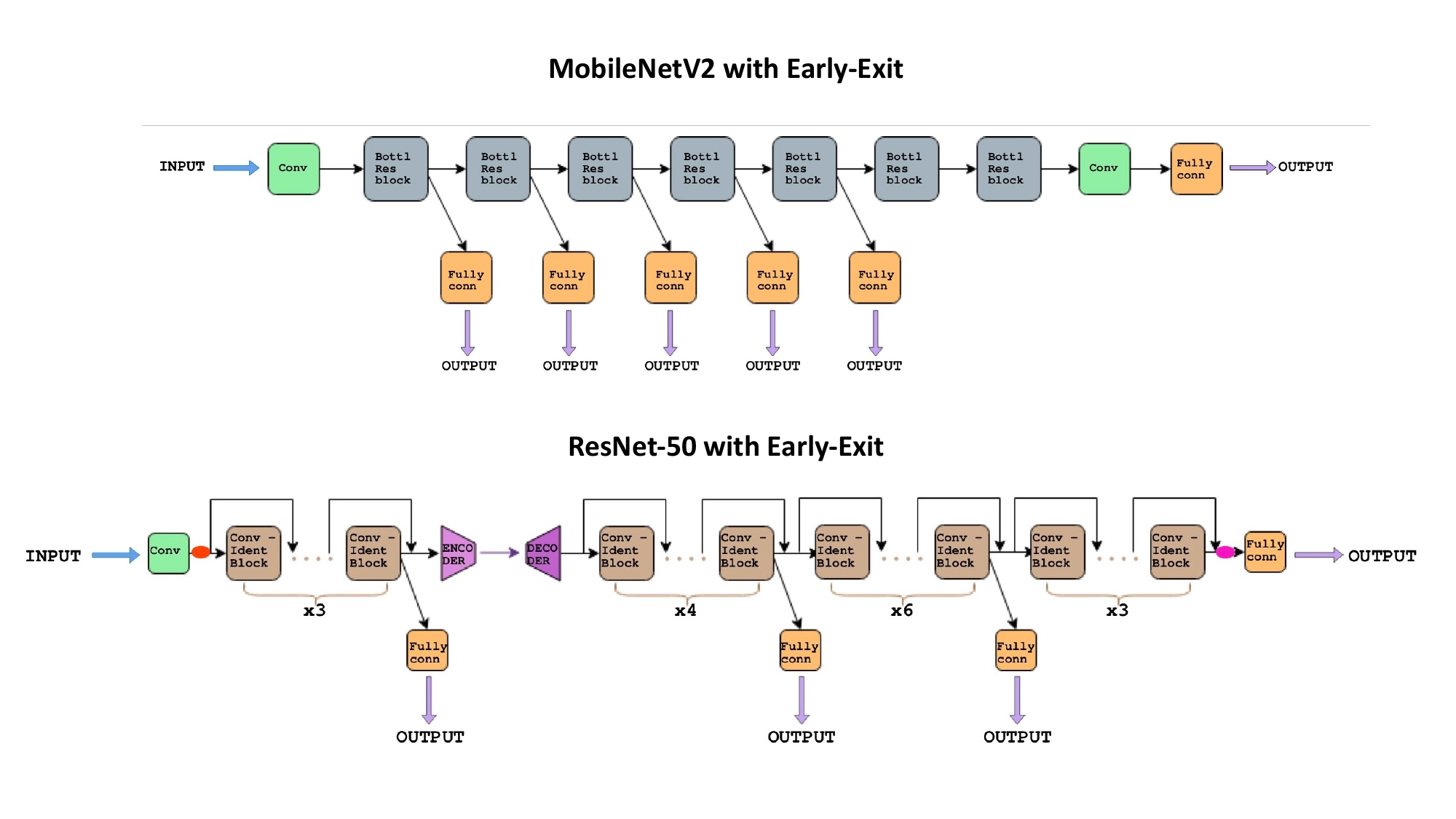}
    \vspace{-20pt}
    \caption{MobileNetV2 and ResNet50 architectures with early-exit points.}
    \label{fig:modelsEE}
    \vspace{-15pt}
\end{figure}

\section{Experimental Results}
\label{sec:results}

In this section, we evaluate the performance of our algorithm MDI-Exit in a testbed consisting of NVIDIA Jetson Nano devices. First, we describe the testbed architecture, DNN models, and the datasets. We then provide the corresponding experimental results.

\textbf{Testbed.} To conduct our experiments, we utilized the NVIDIA\textsuperscript{®} Jetson Nano™, a commonly used platform for AI development on edge devices. The Jetsons are equipped with the NVIDIA Maxwell architecture, featuring 128 NVIDIA CUDA\textsuperscript{®} cores, a Quad-core ARM Cortex-A57 MPCore processor, and 4 GB of 64-bit LPDDR4 memory operating at 1600MHz with a bandwidth of 25.6 GB/s. The Jetsons are connected using WiFi. 

We consider the following topologies in our experiments: (i) ``2-Node'' topology comprised of two workers. (ii) ``3-Node-Mesh'' topology, where three workers are connected to each other in a fully connected manner. (iii) ``3-Node-Circular'' topology which consists of three nodes connected in a circular manner. (iv) ``5-Node-Mesh'' topology of five fully connected workers. We assume that $T_{Q1}=10$, $T_{Q1}=30$, $T_{O} = 50$, $\alpha = 0.2$, $\beta=0.1$, $\zeta=0.2$, $T_e^{\min}$.

\textbf{Datasets and DNN Models.} We use CIFAR-10 dataset \cite{cifar_dataset} utilizing 10,000 test images.  We use the \mobilenetvtwo{} \cite{sandler2018mobilenetv2} and ResNet-50 as DNN inference models. We created five early-exit points in \mobilenetvtwo{}  and three exit points in ResNet-50 as shown in Fig. \ref{fig:modelsEE}. We note that we implemented an auto-encoder after the first exit point in ResNet-50 to reduce the size of the feature vector. We will discuss the impact of using auto-encoder on the performance later in this section. Our auto-encoder architecture is comprised of two convolution layers, where each convolution layer is followed by a RELU function. The autoencoder reduces the feature vector size from 3.2MB to 13.3KB, enabling significant compression and reducing the delay due to transmitting activation vectors. We use the autoencoder only at the first exit point of ResNet-50, because the size of feature vectors in other exit points and MobileNetV2 is smaller, and autoencoder reduces the overall accuracy (up to 2.2\% at the first exit point of ResNet-50).

\begin{figure}[t!]
%\vspace{-5pt}
    \centering
    \scalebox{0.7}{
    \includegraphics{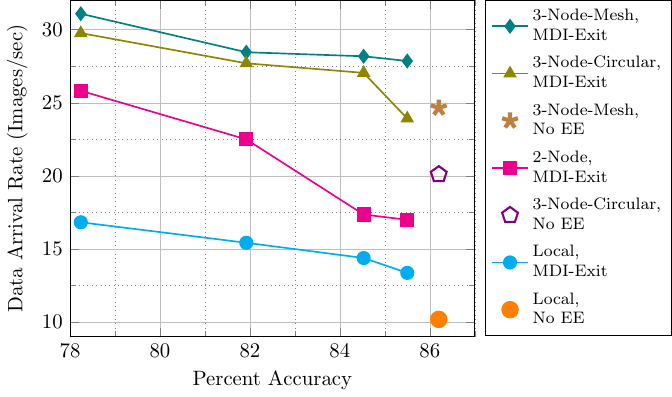}}
    %\vspace{-5pt}
    \caption{MobileNetV2. Early-exit confidence threshold (accuracy) is fixed.}
    \label{fig:mobile-vary-fig}
    \vspace{-5pt}
\end{figure}

\begin{figure}[t!]
%\vspace{-5pt}
    \centering
    \scalebox{0.7}{
    \includegraphics{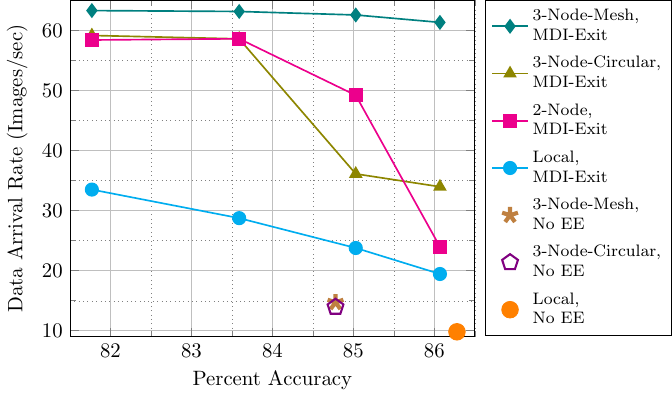}}
    %\vspace{-5pt}
    \caption{ResNet50. Early-exit confidence threshold (accuracy) is fixed.}
    \label{fig:resnet-vary-fig}
    \vspace{-15pt}
\end{figure}

 \textbf{Results.} We first consider the scenario in which the early-exit confidence threshold $T_e^k$ is fixed for all tasks, and MDI-Exit adaptively arranges the data arrival rate. The results are shown in Figs. \ref{fig:mobile-vary-fig} and \ref{fig:resnet-vary-fig} for MobileNetV2 and ResNet-50, respectively. We note that ``Local, MDI-Exit'' means that early-exit happens; the DNN model is processed locally, but the model is not distributed across multiple devices. Also, the points ``3-Node-Mesh, No EE'', ``3-Node-Circular, No EE'', and ``Local, No EE'' correspond to the experiments when there is no early-exit. As seen, MDI-Exit enjoys the tradeoff between data arrival rate and inference accuracy thanks to (i) employing both model-distributed inference and early-exit, and (ii) adaptively deciding when to exit early. Also, when the number of nodes increases, we observe that MDI-Exit achieves a higher data arrival rate thanks to utilizing the computing resources of multiple devices in parallel. Fig. \ref{fig:resnet-vary-fig} confirms the same observations for ResNet-50.

\begin{figure}[t!]
%\vspace{-5pt}
    \centering
    \scalebox{0.9}{
    \includegraphics{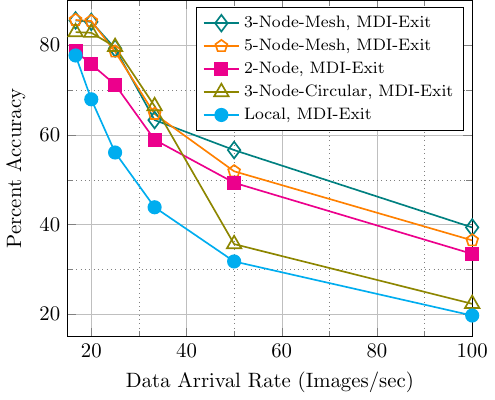}}
    %\vspace{-5pt}
    \caption{MobileNetV2. Poisson arrival with a fixed average arrival rate. }
    \label{fig:mobile-poisson-fig}
    \vspace{-5pt}
\end{figure}

\begin{figure}[t!]
%\vspace{-5pt}
    \centering
    \scalebox{0.9}{
    \includegraphics{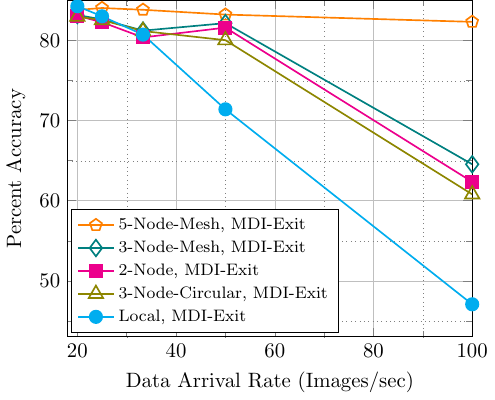}}
    %\vspace{-5pt}
    \caption{ResNet50. Poisson arrival with a fixed average arrival rate.}
    \label{fig:resnet-poisson-fig}
    \vspace{-15pt}
\end{figure}

Next, we consider the scenario that the data arrival rate follows Poisson distribution and the average rate is fixed, and MDI-Exit arranges the early-exit threshold (hence accuracy) to admit all the arriving traffic. Fig. \ref{fig:mobile-poisson-fig} shows the inference accuracy versus the average data arrival rate graph for different topologies. As seen inference accuracy can be kept high even though the data arrival rate is high in multi-node setups thanks to model-distributed inference and task parallelization. We note that MDI-Exit performs better in  3-Node-Mesh topology than 5-Node-Mesh topology because data transmission is too much and becomes a bottleneck in 5-Node-Mesh topology. To address this issue, we used an autoencoder for ResNet-50 as described earlier in this section.  Fig. \ref{fig:resnet-poisson-fig} shows inference accuracy versus data arrival rate results for ResNet-50. As seen, MDI-Exit performs the best in 5-Node-Mesh topology; the inference accuracy slightly reduces with increasing data rate thanks to (i) using autoencoder, which reduces unnecessary transmissions, (ii) model-distributed inference, which allows utilizing the computing power all devices in parallel, and (iii) early-exit, which allows exiting early for some data (images), which saves computing power and transmission bandwidth for more difficult data.

\section{Conclusion}
\label{sec:conc}

In this paper, we investigated the design and development of MDI with early exit, where (i) a DNN model is distributed across multiple workers, and (ii) we can exit the model without processing all the layers if target accuracy is reached. We designed a framework MDI-Exit which adaptively determines early-exit and offloading policies as well as data admission at the source.  Experimental results on a real-life testbed of NVIDIA Nano edge devices showed that MDI-Exit processes more data when accuracy is fixed and results in higher accuracy when the data rate is fixed.

\bibliographystyle{IEEEtran}
%\bibliography{strings,refs}
\bibliography{refs}

% Generated by IEEEtran.bst, version: 1.14 (2015/08/26)
\begin{thebibliography}{10}
\providecommand{\url}[1]{#1}
\csname url@samestyle\endcsname
\providecommand{\newblock}{\relax}
\providecommand{\bibinfo}[2]{#2}
\providecommand{\BIBentrySTDinterwordspacing}{\spaceskip=0pt\relax}
\providecommand{\BIBentryALTinterwordstretchfactor}{4}
\providecommand{\BIBentryALTinterwordspacing}{\spaceskip=\fontdimen2\font plus
\BIBentryALTinterwordstretchfactor\fontdimen3\font minus \fontdimen4\font\relax}
\providecommand{\BIBforeignlanguage}[2]{{%
\expandafter\ifx\csname l@#1\endcsname\relax
\typeout{** WARNING: IEEEtran.bst: No hyphenation pattern has been}%
\typeout{** loaded for the language `#1'. Using the pattern for}%
\typeout{** the default language instead.}%
\else
\language=\csname l@#1\endcsname
\fi
#2}}
\providecommand{\BIBdecl}{\relax}
\BIBdecl

\bibitem{AR-MDI}
P.~Li, E.~Koyuncu, and H.~Seferoglu, ``Adaptive and resilient model-distributed inference in edge computing systems,'' \emph{IEEE Open Journal of the Communications Society}, 2023.

\bibitem{li2023model}
P.~Li, H.~Seferoglu, and E.~Koyuncu, ``Model-distributed inference in multi-source edge networks,'' in \emph{2023 IEEE International Conference on Acoustics, Speech, and Signal Processing Workshops (ICASSPW)}.\hskip 1em plus 0.5em minus 0.4em\relax IEEE, 2023, pp. 1--5.

\bibitem{shallowdeep}
Y.~Kaya, S.~Hong, and T.~Dumitras, ``Shallow-deep networks: Understanding and mitigating network overthinking,'' in \emph{International Conference on Machine Learning}.\hskip 1em plus 0.5em minus 0.4em\relax PMLR, 2019, pp. 3301--3310.

\bibitem{branchynet}
S.~Teerapittayanon, B.~McDanel, and H.-T. Kung, ``Branchynet: Fast inference via early exiting from deep neural networks,'' in \emph{2016 23rd International Conference on Pattern Recognition (ICPR)}.\hskip 1em plus 0.5em minus 0.4em\relax IEEE, 2016, pp. 2464--2469.

\bibitem{neely2010stochastic}
M.~J. Neely, ``Stochastic network optimization with application to communication and queueing systems,'' \emph{Synthesis Lectures on Communication Networks}, vol.~3, no.~1, pp. 1--211, 2010.

\bibitem{sandler2018mobilenetv2}
M.~Sandler, A.~Howard, M.~Zhu, A.~Zhmoginov, and L.-C. Chen, ``Mobilenetv2: Inverted residuals and linear bottlenecks,'' in \emph{Proceedings of the IEEE conference on computer vision and pattern recognition}, 2018, pp. 4510--4520.

\bibitem{deng2009imagenet}
J.~Deng, W.~Dong, R.~Socher, L.-J. Li, K.~Li, and L.~Fei-Fei, ``Imagenet: A large-scale hierarchical image database,'' in \emph{2009 IEEE conference on computer vision and pattern recognition}.\hskip 1em plus 0.5em minus 0.4em\relax Ieee, 2009, pp. 248--255.

\bibitem{teerapittayanon2017distributed}
S.~Teerapittayanon, B.~McDanel, and H.-T. Kung, ``Distributed deep neural networks over the cloud, the edge and end devices,'' in \emph{2017 IEEE 37th international conference on distributed computing systems (ICDCS)}.\hskip 1em plus 0.5em minus 0.4em\relax IEEE, 2017, pp. 328--339.

\bibitem{kang2017neurosurgeon}
Y.~Kang, J.~Hauswald, C.~Gao, A.~Rovinski, T.~Mudge, J.~Mars, and L.~Tang, ``Neurosurgeon: Collaborative intelligence between the cloud and mobile edge,'' \emph{ACM SIGARCH Computer Architecture News}, vol.~45, no.~1, pp. 615--629, 2017.

\bibitem{eshratifar2019jointdnn}
A.~E. Eshratifar, M.~S. Abrishami, and M.~Pedram, ``Jointdnn: an efficient training and inference engine for intelligent mobile cloud computing services,'' \emph{IEEE Transactions on Mobile Computing}, vol.~20, no.~2, pp. 565--576, 2019.

\bibitem{EdgePipe-hu2021pipeline}
Y.~Hu, C.~Imes, X.~Zhao, S.~Kundu, P.~A. Beerel, S.~P. Crago, and J.~P.~N. Walters, ``Pipeline parallelism for inference on heterogeneous edge computing,'' \emph{arXiv preprint arXiv:2110.14895}, 2021.

\bibitem{bolukbasi}
T.~Bolukbasi, J.~Wang, O.~Dekel, and V.~Saligrama, ``Adaptive neural networks for efficient inference,'' in \emph{International Conference on Machine Learning}.\hskip 1em plus 0.5em minus 0.4em\relax PMLR, 2017, pp. 527--536.

\bibitem{biasielli2020neural}
M.~Biasielli, C.~Bolchini, L.~Cassano, E.~Koyuncu, and A.~Miele, ``A neural network based fault management scheme for reliable image processing,'' \emph{IEEE Transactions on Computers}, vol.~69, no.~5, pp. 764--776, 2020.

\bibitem{panda2016cdl}
P.~Panda, A.~Sengupta, and K.~Roy, ``Conditional deep learning for energy-efficient and enhanced pattern recognition,'' in \emph{2016 Design, Automation \& Test in Europe Conference \& Exhibition (DATE)}.\hskip 1em plus 0.5em minus 0.4em\relax IEEE, 2016, pp. 475--480.

\bibitem{koyuncu:c31}
A.~Gormez, V.~Dasari, and E.~Koyuncu, ``{E$^2$CM}: Early exit via class means for efficient supervised and unsupervised learning,'' in \emph{International Joint Conference on Neural Networks (IJCNN)}, Jul. 2022.

\bibitem{han2021dynamic}
Y.~Han, G.~Huang, S.~Song, L.~Yang, H.~Wang, and Y.~Wang, ``Dynamic neural networks: A survey,'' \emph{IEEE Transactions on Pattern Analysis and Machine Intelligence}, 2021.

\bibitem{liu2020pruning}
J.~Liu, S.~Tripathi, U.~Kurup, and M.~Shah, ``Pruning algorithms to accelerate convolutional neural networks for edge applications: A survey,'' \emph{arXiv preprint arXiv:2005.04275}, 2020.

\bibitem{koyuncu:c32}
A.~Gormez and E.~Koyuncu, ``Pruning early exit networks,'' in \emph{Workshop on Sparsity in Neural Networks}, Jul. 2022.

\bibitem{chen2015compressing}
W.~Chen, J.~Wilson, S.~Tyree, K.~Weinberger, and Y.~Chen, ``Compressing neural networks with the hashing trick,'' in \emph{International conference on machine learning}.\hskip 1em plus 0.5em minus 0.4em\relax PMLR, 2015, pp. 2285--2294.

\bibitem{gong2014compressing}
Y.~Gong, L.~Liu, M.~Yang, and L.~Bourdev, ``Compressing deep convolutional networks using vector quantization,'' \emph{arXiv preprint arXiv:1412.6115}, 2014.

\bibitem{wangni2018gradient}
J.~Wangni, J.~Wang, J.~Liu, and T.~Zhang, ``Gradient sparsification for communication-efficient distributed optimization,'' in \emph{Advances in Neural Information Processing Systems}, 2018, pp. 1299--1309.

\bibitem{alistarh2018convergence}
D.~Alistarh, T.~Hoefler, M.~Johansson, N.~Konstantinov, S.~Khirirat, and C.~Renggli, ``The convergence of sparsified gradient methods,'' in \emph{Advances in Neural Information Processing Systems}, 2018, pp. 5973--5983.

\bibitem{cifar_dataset}
\BIBentryALTinterwordspacing
``Cifar dataset.'' [Online]. Available: \url{https://www.cs.toronto.edu/~kriz/cifar.html}
\BIBentrySTDinterwordspacing

\end{thebibliography}

% \iffalse
% \newpage
% \input{Appendix}
% \fi

\end{document}